# Collective effects in charge transfer within a hybrid organic-inorganic system

Y. Paltiel,[1*] G. Jung,[2] T. Aqua,[3] D. Mocatta,[4] U. Banin,[4] and R. Naaman[3*]

[1] Applied Physics Department and the Center for Nano Science and Nanotechnology, The Hebrew University, Jerusalem 91904, Israel

[2] Department of Physics, Ben Gurion University of the Negev, 84105 Beer-Sheva, Israel

[3] Department of Chemical Physics, The Weizmann Institute, Rehovot 76100, Israel

[4] Institute of Chemistry and the Center for Nanoscience and Nanotechnology, The Hebrew University of Jerusalem, Jerusalem 91904, Israel

## Abstract

A collective electron transfer (ET) process was discovered by studying the current noise in a field effect transistor with light-sensitive gate formed by nanocrystals linked by organic molecules to its surface. Fluctuations in the ET through the organic linker are reflected in the fluctuations of the transistor conductivity. The current noise has an avalanche character. Critical exponents obtained from the noise power spectra, avalanche distributions, and the dependence of the average avalanche size on avalanche duration are consistent with each other. A plausible model is proposed for this phenomenon.

Charge transfer through organic molecular systems was extensively studied both theoretically and experimentally.[1,2,3] In many biologically related systems, the electron transfer (ET) occurs within supramolecular two-dimensional structures, such as membranes. These membranes are constantly fluctuating due to thermal motion. Relation between these fluctuations and the rate of charge transfer was found in some cases.[4] The possibility of cooperative effects in charge transfer has been raised before both in relation to biological systems[5] and clusters of polar-polarizable chromophores.[6,7] The effect was attributed, in some systems, to photo-induced phase transition in the crystals.

In the present work we employ noise techniques to investigate the dynamics of charge transfer between semiconductor nanocrystals and a semiconductor surface linked by a self-assembled organic monolayer. While usually one studies the ET events as if they are not correlated with each other, here we present evidences that structural fluctuations may lead to avalanche of ET events. We find correlations between ET events that lead to the propagation of charge avalanches. The avalanche does not depend on the light intensity and therefore the present findings may be relevant to biological systems, where naturally the photo induced events occur at relatively low illumination.

The method applied and the system studied here offer a unique model for investigating dynamics of multiple ET events in two-dimensional supramolecular structures.

The experimental arrangement, shown in Fig. 1A (inset), consists of InAs nanocrystals (NCs), linked by a self-assembled organic monolayer (SAM) to the surface of a two-dimensional electron gas (2DEG) GaAs/AlGaAs field effect transistor (FET). The system was shown to operate as a sensitive infrared photodetector with the NCs acting as light-sensitive gates for the FET.[8] When the NCs absorb light, the photo-excited electrons are transferred to the FET surface states (traps) via the organic molecules, change the surface potential, and thus modify the conductivity of the 2DEG transistor channel. After some characteristic life-time in the trap, the electrons are transferred back to the NCs and recombined with photoexcited holes. The FET operates not only as a light detector but also as an amplifier of the light effect. A change in the FET current can be orders of magnitude larger than the number of electrons transferred through the molecular self-assembled layer.[9]

In the present study, the InAs NCs are illuminated with photons of energy below the band gap of the GaAs, preventing the FET device itself to be directly excited by light. Therefore, the response of the FET conductivity provides direct and amplified signatures of fluctuations in the charge transfer through the self-assembled organic molecular structure.

The fabrication process of NC-gated FET has been described elsewhere.[9] In brief, first a gateless GaAs FET is fabricated. The conducting channel in the structure is confined to the n-GaAs layer at a depth of 20 nm from the surface. A self-assembled organic monolayer of 1,9 nonanedithiol [HS-$(CH_2)_9$-SH] is deposited between the source and drain electrodes.[10] Finally, the monodisperse InAs NCs (dia. of 6 nm, 0.85 eV exciton gap) are attached to the organic monolayer.

The dynamics of charge transfer was investigated by measuring, in the frequency and time domain, fluctuations of the conductivity in the FET, biased with a constant voltage source. The resulting current noise provides an insight into the dynamics of charge transfer through the organic SAM. The voltage biasing circuit contains large ballast resistors connected in series with the sample, and the noise current signal is enhanced by a very low noise transimpedance preamplifier, both of which minimize the effects of contacts and fluctuations due to the measuring chain and environment.

Time traces of current fluctuations recorded at room temperature under dark conditions and under infrared (IR) illumination are shown in Fig. 1B. The dark noise appears as featureless fluctuations with a Gaussian distribution of amplitudes. IR light illumination leads to the appearance of current spikes in time domain records and to significant deviations from Gaussianity. Current spikes above and below the mean current level reflect abrupt increases and decreases in the conductivity, respectively. Zoom on the spikes (inset Fig. 1B) reveals avalanche-like events. Since no spikes were observed in time traces recorded with reference sample in which no NCs were deposited on top of the SAM, we conclude that current avalanches are solely associated with the light-induced charge transfer processes. Moreover, at a given voltage bias, the height of the largest current spikes is of the order of the dc response to the light illumination, consistent with the prediction that the spikes with maximum height appear when the entire surface of the FET channel is gated. Intensity dependent studies of the signal did not reveal any non-linearity.

Avalanche-like noise arises when a system responds to changing external conditions through discrete, impulsive events spanning a broad range of sizes.[11] A wide variety of physical systems exhibiting avalanche noise have been studied. These systems exhibit collective behavior characterized by the properties of the changing avalanche parameters. A common feature of systems with avalanche dynamics is that statistics of various avalanche parameters, such as height, size, or duration, lack a characteristic scale.[11] The scale invariance manifests itself in power law distributions of avalanche parameters. Indeed, the size of the detected current avalanches (S) is power law distributed, $S \equiv \int I(t)dt \propto S^{\tau}$ , as shown in Fig. 2A. The best fit (solid line) to a power law obtained with the exponent $\tau = 2.02 \pm 0.08$. Typically, the size distribution exponents are smaller than 2 and cluster around 1.5 or 1.7. However, exponents higher than 2 have been found before.[Error! Bookmark not defined.,12] Deviations of the distribution from the power law, seen in Fig. 2A at the high sizes edge, are caused by large avalanches spanning the entire sample size.[12] Flattening of the distribution at the low size end is due to plunging of small avalanches into the Gaussian background noise.

The avalanche duration and height were also found to be scale invariant. The avalanche durations are distributed according to the power law $T \propto T^{\alpha}$, with $\alpha = 2.2 \pm 0.2$, while an exponent $2.7 \pm 0.2$ provides the best fit of the height distribution to the respective power law. An avalanche height distribution exponent of 2.7 was reported to characterize systems with avalanches following the binary tree structure.[13] This suggests that avalanches in our system may be propagating along the same pattern.

Figure 2B shows the power spectrum of the current noise characterizing only fluctuations in charge transfer through the organic linker. The spectrum was obtained by subtracting the background noise spectrum, measured under dark conditions, from the data measured under IR illumination. The noise scales as the power law of frequency and exhibits two different power law exponents. At high frequencies, the spectrum exponent $1.0 \pm 0.1$ reflects the dynamics within avalanches, whereas the low-frequency, exponent of $3.1 \pm 0.1$ reflects correlations between avalanches. The latter is consistent with the observed power law distribution of waiting times between avalanches.

Scale invariance observed in many systems with avalanche dynamics is often attributed to the proximity of a non-equilibrium phase transition or to the self-organization of a system to a critical point which is an attractor of the dynamics. The exponents of power law distributions in critical avalanche dynamics are universal and independent of experiments within a given universality class. They are related by the critical exponent relation, $\alpha - 1 = (\tau - 1)/\sigma\nu z$.**Error! Bookmark not defined.** The product σνz determines the exponent of the noise spectral density, which is expected to scale at high frequencies as $S(\omega) \propto \omega^{1/\sigma\nu z}$ and appears also as the exponent of power law that characterizes the average avalanche size <S> dependence on the avalanche duration $\langle S \rangle \propto T^{1/\sigma\nu z}$.**Error! Bookmark not defined.**

The dependence of the average avalanche size on the avalanche duration is shown in Fig. 2C. The exponent $\sigma\nu z = 1.1 \pm 0.1$, obtained by fitting the average avalanche size to the power of the duration, agrees well with the exponent obtained by fitting the high frequency part of the current noise spectra to the power law $S(\omega) \propto \omega^{1/\sigma\nu z}$. Moreover, the experimentally determined exponents fulfill the critical exponents' relation $\alpha - 1 = (\tau - 1)/\sigma\nu z$.

The noise data indicate that charge transfer through self-assembled organic molecules occurs as correlated, uninterrupted sequences of events, avalanches, rather than acts of individual electron transfer. To trigger an avalanche, a perturbation facilitating initial charge transfer needs to occur. Such a perturbation can cause a burst of activity by triggering subsequent charge transfer events in its vicinity. In a critical system, the maximum size of such a chain reaction is limited only by the finite system size. The self-assembled organic molecules serve as bridges with energy barriers for charge transfer between the NCs (donors) and the surface states (acceptors) localized in the oxide layer underneath the bound molecules. Since avalanches occur above and below the average dc current level (Fig. 1B), we observe charge transfer events occurring both from the NCs to the surface states and back from the surface states to the NCs.

In principle, charge avalanches can be triggered either by a decrease in the barrier height for the ET, resulting from a structural change occurring in organic molecules, or by a shift in the relative energies of the acceptor and donor states due to the transfer of the first electron, which facilitates sequential transfer of other photoexcited charges. In the following, we discuss the two plausible mechanisms. Based on this discussion and on additional experiments, we suggest that one of the mechanisms is more probable.

Let us consider the first model that assumes structure induced avalanche. Nonaedithiol molecules in the SAM may have either an all-trans configuration or some of the bonds in the molecules may have the gauche structure (Fig. 3B). At high temperatures, the gauche configuration prevails, whereas at low temperatures the all-trans configuration dominates.[14] At room temperatures there is a finite probability of spontaneous, thermally activated transitions between both configurations.[15,16] The gauche state is characterized by a high-energy barrier for ET, which reduces the charge transfer probability, whereas the barrier in the all-trans is relatively low and allows for efficient charge transfer across the bridge (Fig. 3A).[14] The surface state in the GaAs side is a deep impurity state which accounts for the asymmetric well behavior in both charge transfer directions.[17] Hence, the trigger that initiates an avalanche is a spontaneous transition of a molecule, serving as a bridge, from a gauche to the all-trans state. Further avalanche propagation is sustained by subsequent modification of neighboring molecular linkers, as a result of the coupling between the molecules, which results in minimizing the repulsion between the molecules by synchronized transfer of many neighboring molecules to the all-trans configuration (see Fig. 3B).

The second plausible model is related to the energy gap between the donor and acceptor states. In the system studied, the donor-acceptor energy gap is consistent with ET in the “inverted region”.[18] The Marcus-theory[19] predicts that if the difference in free energy, $\Delta G$, between the donor and acceptor states is large and the reorganization energy, $\lambda$, of the system is smaller than the free energy, then a decrease in $|\Delta G|$ will result in an increase in the rate of electron transfer.[20] These conditions are referred to as the “inverted regime”.  In our system, Coulomb interaction between adjacent transferred electrons could, in principle, create the conditions for a collective effect in which the first electrons that is transferred reduces $\Delta G$ and therefore the transfer of additional electrons is enhanced.

In order to determine which of the above two mechanisms is valid, we have conducted temperature-dependent studies. The room temperature critical avalanche dynamics causes significant deviations from Gaussianity at high amplitudes (Fig. 4). At low temperatures, the fluctuations of the transistor current are purely Gaussian (Fig, 4 inset). This temperature effect can be rationalized based on the first model. At low temperatures, most of the nonaedithiol molecules are in the all-trans state and the probability of a transition into the gauche configuration is very low.[14] Therefore, the charge transfer probability does not vary with time and no avalanches develop. However, at room temperature, fluctuations in the molecular structure take place and avalanche charge transfer dynamics appears. The apparent critical behavior of the system can be attributed to the operation in the vicinity of structural order (all-trans)-disorder (gauche) transition in the molecular bridges. The second model which is based on the reduction of the energy gap between the donor and acceptor states, is not expected to result in eliminating the avalanche behavior at low temperatures. In addition, in the NCs, the initially excited 1p state decays very rapidly to the 1s state (sub picoseconds timescale)[21] and therefore the large energy gap, that may support the avalanche according to the second model, does not exist long enough to allow for the charge transfer process to occur. We conclude therefore, that the most probable mechanism is that the avalanches are triggered by fluctuations in the cooperative structural transitions of the organic molecules serving as the bridge in the ET process.

Discrete avalanche events in many systems result from applying a slow, continuously varying, external perturbation. Here, the system crackles under continuous excitation by light illumination. Interactions between the molecules and donor states are responsible for time correlations between avalanches and power law distribution of waiting times, while phase transition between the ordered and disordered molecular structural state lies behind the critical behavior of our system.

In summary, we have employed current noise measurements in a FET coupled to an NC system in order to probe the charge transfer processes through the molecules in a self-assembled molecular layer. The noise seen, under infrared illumination, has an avalanche character, which indicates that the electron transfer is a cooperative effect. We suggest that structural process lies behind the observed avalanches. This type of structural transitions take place also in biological membranes and may affect multiple ET events also there.

The system described here can be modified so as to tune its various parameters. Modification can be induced by varying the organic molecules that couples the nano crystals to the semiconductor device and varying the various parameters of the conducting channel in the solid state device.

**Figure captions:**

**Figure 1:** A) SEM micrograph of the InAs nanocrystals (NCs) layer. Inset shows schematics of the FET and NCs arrangements. B) Room temperature time traces of current fluctuations in an AC-coupled FET channel at a low bias of 0.04V under 1060 nm IR illumination of 10 mW in a spot of 1mm diameter (upper panel) and under dark conditions.

**Figure 2: A).** Distribution of the current avalanche sizes in Ampere·sec units. The dashed line represents the best fit to the power law in the intermediate size range. B). Power spectral density of noise induced by IR light illumination. Solid lines represent the best fits to the power law. C). Average avalanche size <S> as a function of avalanche duration T, with the best fit to the power law $\langle S\rangle \propto T^{1/\sigma\nu z}$ (dashed lines).

**Figure 3:** Schematics of the mechanisms for charge avalanche triggering and propagation. A) The avalanches are triggered by a structural transition in the molecules and propagate due to coupling between the molecules, which enforces transitions to all-trans state, thus modifying the barrier height indicated by the bold dash lines in the bridge region. The surface state in the GaAs side, indicated by the bold dash-dot line, is a deep impurity state which account for the asymmetric well behavior in both charge transfer directions. Valence and conduction band levels of the InAs dot are indicated by the dash-dotted bold lines in the NC region. B) Schematics of the molecular layer structure showing all-trans domains and molecules in the gauche configuration.

**Figure 4:** Distribution of the FET current fluctuations in the illuminated device at 300K where a clear non-Gaussian distribution is observed at high amplitudes. The inset shows the Gaussian distribution of current fluctuations at 80 K.

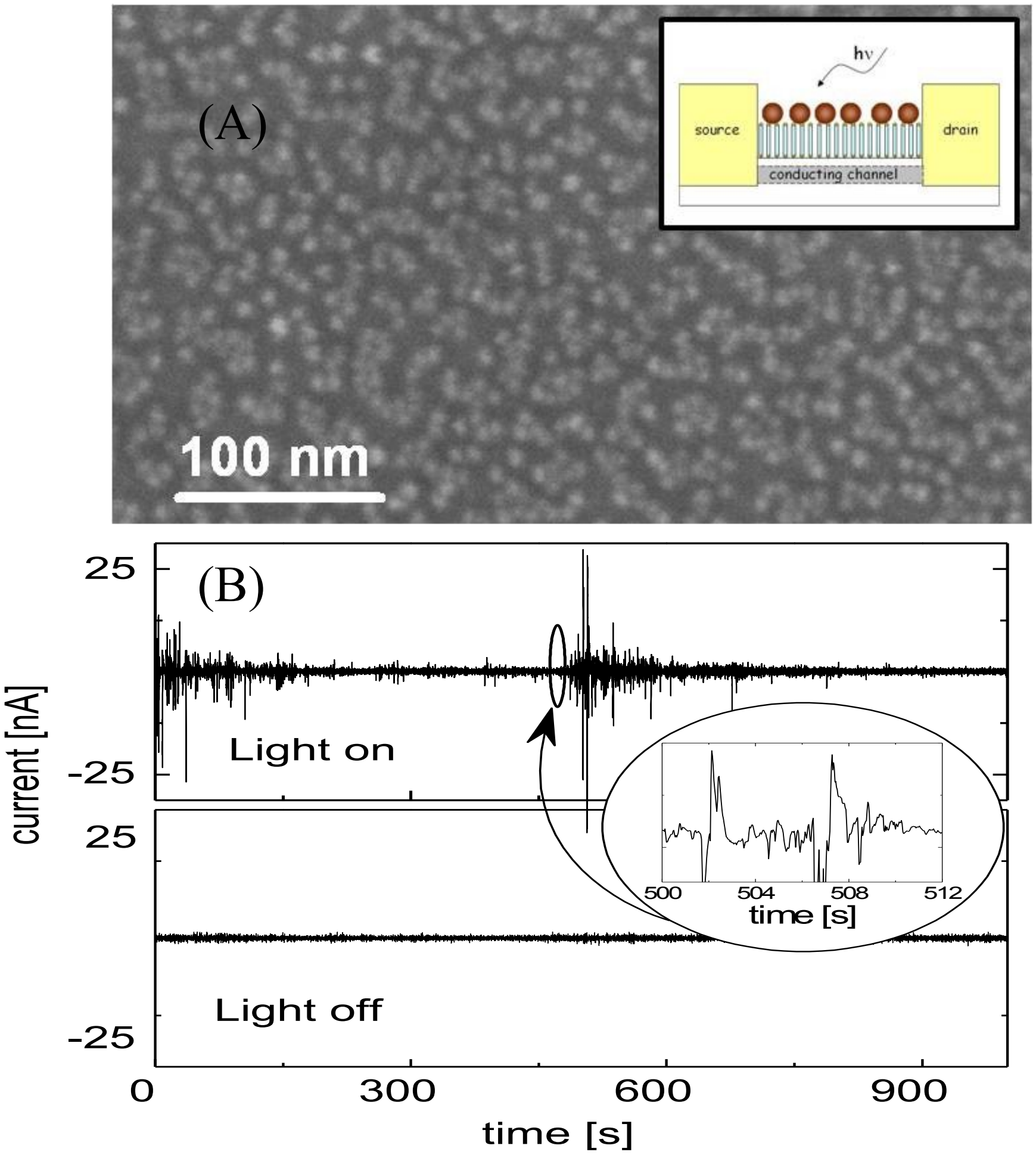
(A)
hν
source
drain
conducting channel
100 nm
(B)
25
-25
current [nA]
Light on
Light off
500
504
508
512
time [s]
0
300
600
900
time [s]

**Figure 1**

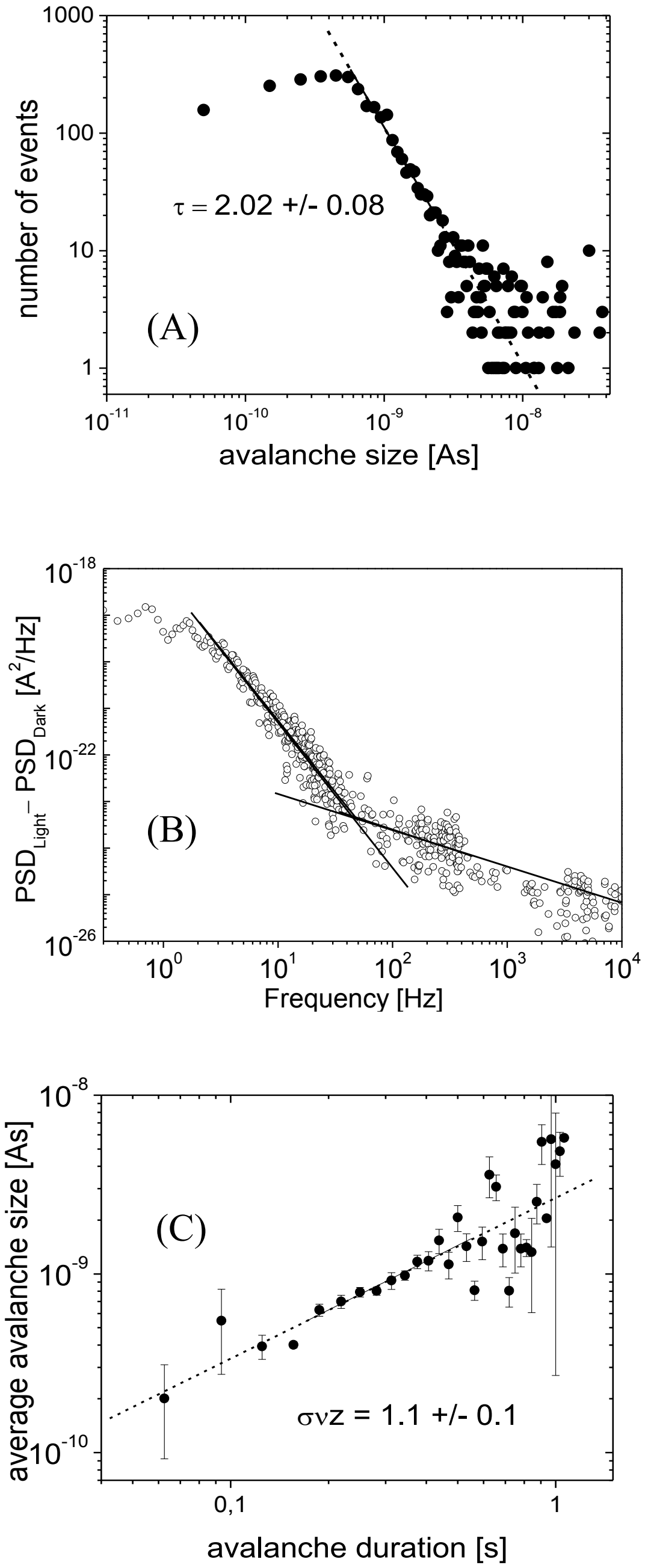
number of events
avalanche size [As]
τ = 2.02 +/- 0.08
(A)
$PSD_{Light} - PSD_{Dark}$ [$A^2$/Hz]
Frequency [Hz]
(B)
average avalanche size [As]
avalanche duration [s]
(C)
σνz = 1.1 +/- 0.1

**Figure 2**

**Figure 3**

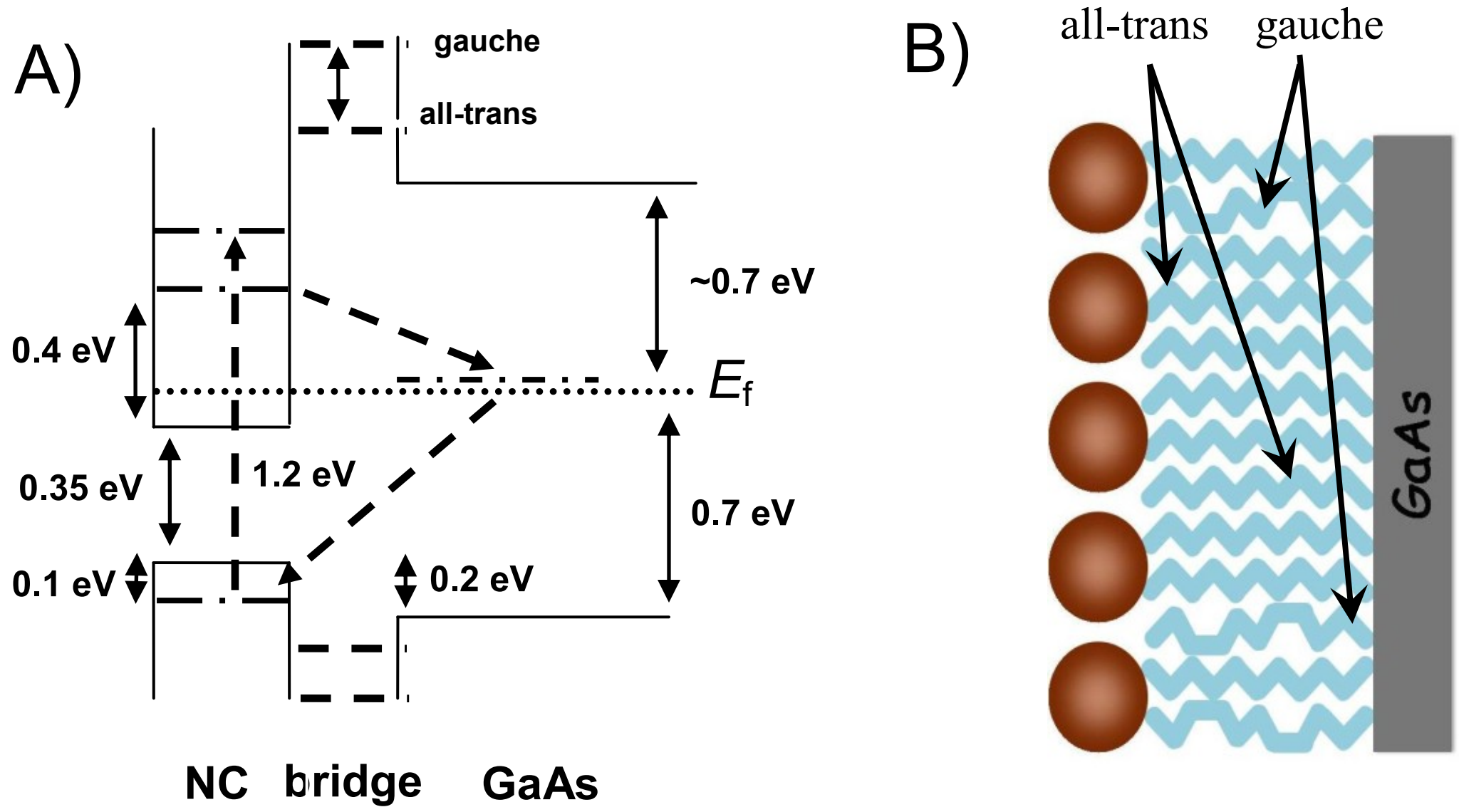
A)
gauche
all-trans
~0.7 eV
0.4 eV
$E_f$
0.35 eV
1.2 eV
0.7 eV
0.1 eV
0.2 eV
NC
bridge
GaAs
B)
all-trans
gauche
GaAs

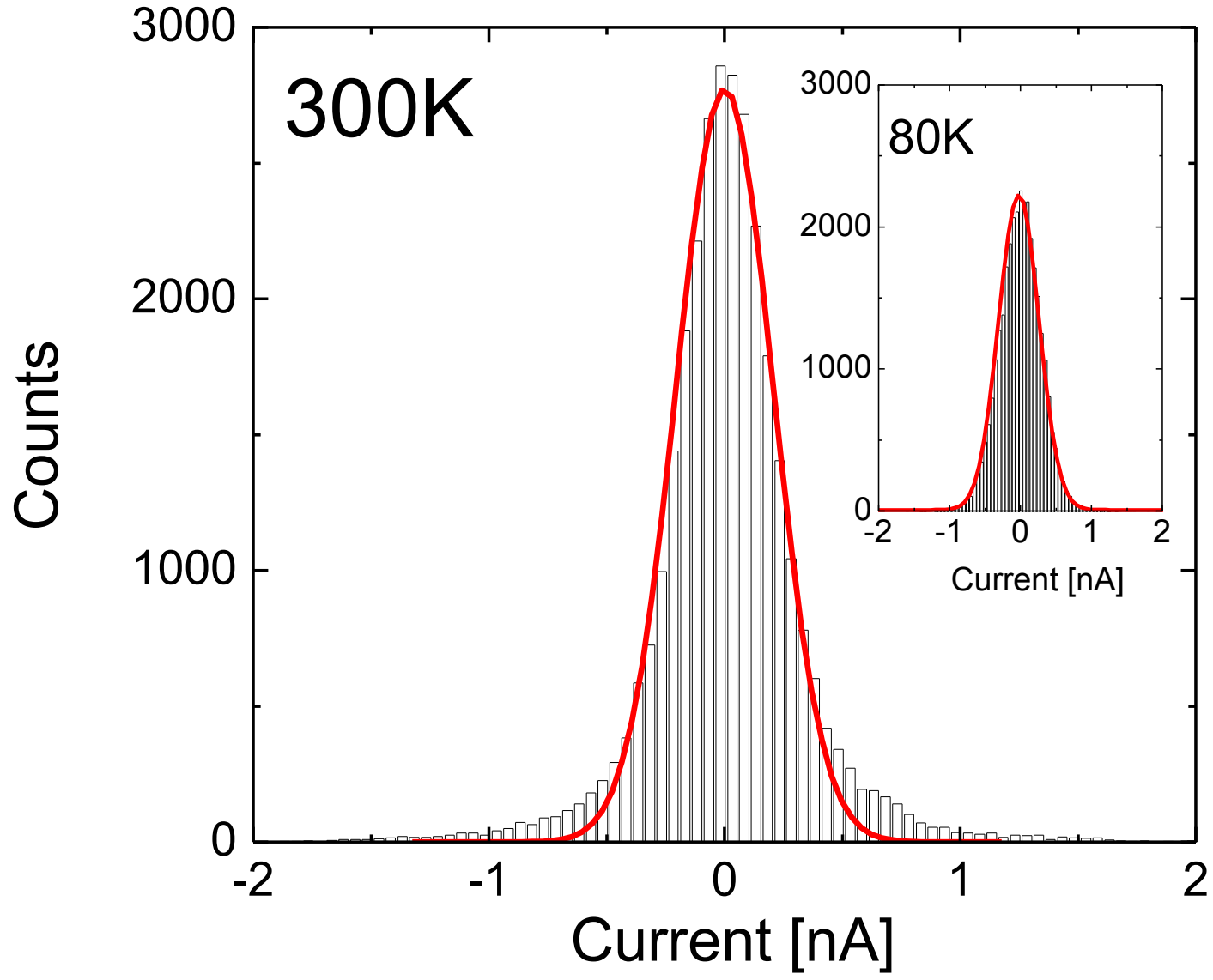

**Figure 4**

**Acknowledgements:** This work was partially supported by the Israeli Ministry of Science grant No. 3444 and by the Grand Center at the Weizmann Institute.